# Anomalous dimensionality dependence of diffusion in a rugged energy landscape: How pathological is one dimension ?

Kazuhiko Seki,[1] **Kaushik Bagchi**[2] and **Biman Bagchi**[3 a)]

[1] *NMRI, National Institute of Advanced Industrial Science and Technology (AIST), Tsukuba Higashi 1-1-1, 305-8565, Japan*

[2] *Department of Mathematics, Ohio State University, Columbus, Ohio 43201, and Chitrakut Annexe, 4th main Road, Malleswaram, Bangalore 560012, India.\**

[3] *Institute of Molecular Science, Okazaki, Aichi, 444 Japan and , Solid State and Structural Chemistry Unit, Indian Institute of Science, Bangalore – 560012, India[*]*

## Abstract

Diffusion in one dimensional rugged energy landscape (REL), is predicted to be *pathologically different (from any higher dimension) with much larger chance of encountering broken ergodicity* (D. L. Stein and C. M. Newman, AIP Conf. Proc. 1479, 620 (2012)). However, no quantitative study of this difference has been reported, despite prevalence of multidimensional physical models in literature (like a high dimensional funnel guiding protein folding/unfolding). Paradoxically, some theoretical studies of these phenomena still employ a one dimensional diffusion description for analytical tractability. We explore the dimensionality dependent diffusion on REL by carrying out an effective medium approximation based analytical calculations and compare them with the available computer simulation results. We find that at intermediate level of ruggedness (assumed to have a Gaussian distribution), where diffusion is well-defined, *the value of the effective diffusion coefficient depends on dimensionality and changes (increases) by several factors (~5-10) in going from 1d to 2d. In contrast, the changes in subsequent transitions (like 2d to 3d and 3d to 4d and so on) are far more modest, of the order of 10-20% only.* When ruggedness is given by random traps with an exponential distribution of barrier heights, the mean square displacement is sub-diffusive (a well-known result), but the growth of MSD is described by different exponents in one and higher dimensions. The reason for such strong ruggedness induced retardation in the case of one dimensional REL is discussed. We also discuss the special limiting case of infinite dimension (d=∞) where the effective medium approximation becomes exact and where theoretical results become simple. We discuss, for the first time, the role of spatial correlation in the landscape on diffusion of a random walker.

[a] Corresponding author email: profbiman@gmail.com;   *) Present address

# I. Introduction

Rugged energy landscape (REL) with spatially distributed maxima and minima is often employed in applications of physics, chemistry and biology (enzyme kinetics, protein DNA interaction and protein folding, diffusion in disordered solids, transport in organic semiconductors, relaxation in random spin systems, in supercooled liquids and glasses). Diffusion of a tagged particle on a complex energy landscape has a long history [1-38]. Such models were initially developed to explore the effects of random disorder on the electrical conductivity, essentially to study the retardation of diffusive motion of electrons and quasi-particles in solids due to the in-built disorders arising from the presence of impurity [4-10]. Several different stochastic models were introduced to account for the presence of random barriers and traps that retard the rate of migration of electrons. In an influential early paper, Scher and Lax introduced the use of the formalism of waiting time distribution to explain the observed power law decay of current in disordered materials [4-6]. Kehr and Haus [11-13] used "hopping over barriers" model to take into account randomly placed barriers and traps. Subsequently, many different theoretical studies were carried out in different areas of condensed matter physics and chemistry to include effects of complex environments on diffusion. However, use of rugged energy landscape gained popularity with the advent of random Ising model which was develoed to address rather unusual properties of spin glasses, such as generation of metastability and slow dynamics. [1,39-41] In the context of transport in the supercooled liquid, Goldstein and Johari discussed the possible influence of energy landscape in describing slow relaxation in glassy liquids. [16] This pioneering idea was further studied using inherent structure formalism [17-23]. The latter provides evidence of the migration of a system from one minimum to another minimum. In another important area where the idea of diffusion in a rugged energy landscape has found wide use is protein folding [24-28].

The mapping of many-particle complex dynamics of the system to a rugged energy landscape with given statistical properties, although approximate, offers a simplification that helps capturing the essence of such aspects like temperature dependence and/or material dependence in glassy liquids, disordered solids and protein diffusion or protein-DNA interaction. One hopes that with only a few measures, like ruggedness energy scale and the correlation length, such an approach can illuminate some aspects of the otherwise complex dynamics, and provide a simpler physical description.

While the landscape in all the above examples is intrinsically multidimensional, theoretical discussions almost always employed one dimensional treatment. This is partly because dimensionality of diffusion is usually considered via Einstein's definition in the following fashion [42]

$$D(d) = \lim_{t \to \infty} \frac{\langle (\Delta r)^2 \rangle}{2dt} \quad , \tag{1}$$

where $\langle (\Delta r)^2 \rangle$ is the mean square displacement of a tagged particle, *t* is the time and *d* is the dimension. The above definition, valid for an ergodic system, is the standard starting point of diffusion in *d*-dimensional system, and in most cases dependence is removed by dividing by *d* (as in Eq. 1). Thus, if we consider random walk in a uniform d-dimensional lattice (like simple cubic) without ruggedness of any kind then *the diffusion constant is independent of d.*

Recently, an elegant study potentially of far-reaching consequence has been carried out by Newman and Stein [39-41] who addressed the issue of dimensionality of diffusion in a rugged landscape. These authors treated diffusion as a percolation invasion problem and concluded that diffusion in one dimension is pathological because the particle (in their language "water in a lake or river") can get trapped ("cannot flow to the sea") due to insurmountable barriers on both sides of exit. This conclusion flows from the observation that the height of the barriers encountered by the walker grows with time as *T log t*, where *T* is the temperature. Therefore, as time increases, the height of barriers encountered increases, the probability of walked getting reflected back and retracing the same path (in 1d) increases, leading to a sharp decrease of diffusion constant (even going to zero asymptotically), if we take the limit of time going to infinity at constant temperature for systems with constant ruggedness, and thus raising the possibility of facing *an ergodicity that is broken or at least compromised*. This may not rule out the existence of diffusion constant with a well-defined value at intermediate times. The situation is different in higher dimensions, including 2d, *because the walker can find practically an infinite number of escape routes, so that "the water can flow to the sea"*. In an earlier publication [41], the same authors considered the problem of broken ergodicity in the problem of diffusion in a rugged landscape, and arrived at similar conclusion about possible lack of diffusion in the asymptotic limit. As noted, this issue of broken ergodicity in rugged landscape does not arise in higher dimensions.

The two basic parameters that are used to characterize the energy landscape are the width of the (assumed) Gaussian distribution of energy (as in spin glasses) and the correlation length that describes spatial relationship. In this work we define ruggedness by a Gaussian distribution of energies at lattice sites of a hypercube. The energies can assume either positive or negative values, and can be correlated over space. The definitions and the model employed bear close resemblance to the treatment of Stein and Newman, and of Zwanzig [27,39-41].

In two recently published studies, we have explored the role of ruggedness on diffusion [37,38]. In the first study, we interrogated (for the first time) by computer simulations the quantitative validity of the well-known expression of Zwanzig on the dependence of diffusion on ruggedness [37]. We found that Zwanzig's expression breaks down due to the presence of three site traps formed by one deep minimum flanked by two large maxima on two sides. We presented a correction term that accounts for the simulation results quantitatively. In the second study we explored the relation between diffusion and entropy in a rugged energy landscape. In particular, we presented a statistical mechanical derivation that showed that the Rosenfeld diffusion-entropy scaling can be recovered *exactly* in the rugged energy landscape [34,35,38].

The study presented in this article has a direct bearing on the role of higher (than one) dimension. Rough or rugged energy landscape has been used to explain distribution of relaxation times observed in enzyme kinetics [32].

In an interesting application of dimension dependent diffusion, Slutsky and Mirny [29] suggested that the efficient search by a protein of the binding site on a DNA may involve a combination of one and three dimensional diffusion. The protein may slide along the DNA in a one dimensional diffusion, but can switch over to a three dimensional diffusion when faced with a bottle-neck along its sliding motion. The combined use of one and three dimensional mode of diffusion is expected to reduce the search time in this complex landscape of diffusion. This model was extended to include the effects of a rugged energy landscape to account for the heterogeneity along the DNA chain [30,31]. The *multidimensionality of energy landscape* is also clearly seen on the unfolding dynamics of the small protein chicken villin headpiece (HP-36). [26] Here the rate determining step of unfolding is found to be the opening of the hydrophobic core formed by three phenyl alanines (Phe-7, phe-11 and phe-18). When heated or solvated in DMSO (Dimethyl sulfoxide), this core is found to melt where Phe-18 first breaks away. Next the contact between Phe-7 and Phe-11 is broken. If we attempt to construct the energy landscape of this process, we shall consider the distance separation between Ph-11, Phe-7 and Phe-18 as the three coordinates that define the relevant landscape.

As discussed by Stein and Newman, diffusion in rugged energy landscape is strongly dimensionality dependent with the possibility of broken ergodicity being a serious concern in systems of reduced dimensionality. [39-41] The objective of the present study is thus deeply rooted in realistic problems in a vast majority of disciplines.

As mentioned earlier, we recently carried out an investigation of the relationship between entropy (S) and diffusion (D) in a rugged landscape and established the relation proposed by Rosenfeld a few years back. [34-38] In the course of the work we noticed that there can indeed be certain significant differences in the determination of diffusion in one dimension (1d). We note that entropy-diffusion relation as envisaged by Rosenfeld scaling relation is oblivious to this difference [34-38].

The relation between diffusion and entropy has a long and illustrated history. In addition to Rosenfeld scaling, the relation proposed by Adam and Gibbs finds wide use, particularly in explanation of glassy dynamics. In this relation, diffusion coefficient decreases sharply as an entropy crisis drives configuration entropy to zero near glass transition. An entropy crisis may develop through emergence of ruggedness, and one may envisage a cross-over from Rosenfeld to Adam-Gibbs scenario.

The objective of the present work is to further explore this dimensionality dependence of diffusion in a rugged landscape and the motivation is provided by the inspiring work of Stein and Newman. Our theory and calculations are based on a simple cubic lattice with random site energies that can be both positive and negative, thus creating the ruggedness.

We find that diffusion in one dimension is indeed markedly different from higher dimension. We, however, are not sure whether the difference can be termed pathological because for small to intermediate range of ruggedness, diffusion in 1d can be lower by a factor of ~5 to 10 or even larger (than in 2d and 3d) while the difference between any two consecutive higher dimension, like 2d and 3d, is only 20-30%.

In the subsequent chapters we introduce the models studied, the theoretical analysis of the dimensionality dependence of diffusion in random lattices, compare the results with available simulation results, consider the asymptotic limit of infinite dimension, and present some results on effects of correlation in the energy landscape. We conclude with a discussion of results and future problems.

## II. Quantitative models of rugged energy landscape

In his landmark paper, Zwanzig considered a continuous rugged potential, U(x). $U(x)$ is assumed to be composed of a back ground potential $U_0(x)$ and a rugged potential $U_1(x)$ and we have $U(x)=U_0(x)+U_1(x)$. $U_1(x)$ is random with a Gaussian distribution. The effective diffusion constant ($D_{eff}$) is obtained as [27,28]

$$D_{eff,z} = \frac{D_0}{\langle e^{U_1/(k_BT)} \rangle \langle e^{-U_1/(k_BT)} \rangle}, \qquad (2)$$

where $D_0$ is the diffusion constant on the smooth potential, $\langle \cdots \rangle$ denotes the spatial average of the rugged potential, $k_B$ is the Boltzmann constant, and $T$ is the temperature. The subscript z of $D_{eff,z}$ indicates Zwanzig's expression obtained using a continuous potential. When the amplitude of the rugged potential is given by a Gaussian distribution with a mean zero and variance $\sigma$,

$$P(U_1) = \frac{1}{\sqrt{2\pi\sigma^2}} \exp\left(-\frac{U_1^2}{2\sigma^2}\right) , \qquad (3)$$

Zwanzig expressed the effective diffusion constant in the following elegant form, [27]

$$D_{eff,z} = D_0 \exp\left[-\sigma^2/(k_B T)^2\right] . \qquad (4)$$

Zwanzig's derivation relies heavily on a local average to smooth the rugged potential. To avoid the local averaging and consider the effective diffusion constant in *d*-dimension, we introduce a hyper-cubic lattice of *d*-dimension. The random energy of i-site is assumed to obey the same Gaussian distribution given by $P(U_i)$ of Eq. (3). We consider two models of rugged landscape. In general random energy landscape, the site energy can be both local maximum and minimum as shown in **Fig. 1**(A). We denote the transition rate from i-site to j-site by $\Gamma_{ij}$. The transition rates are given by [37]

$$\Gamma_{ij} = \begin{cases} \Gamma_0 & U_j \leq U_i \\ \Gamma_0 \exp\left(\dfrac{U_i - U_j}{k_B T}\right) & U_j > U_i \end{cases}, \tag{5}$$

which are identical to those known as the Miller-Abraham process. [9,10] In the Metropolis Monte Carlo scheme for constant temperature, the above rate is implemented to generate configurations via the Boltzmann factor. [3]

As a reference, and to emphasize the difference of diffusion in one dimension from any other dimension, we also present the results of trap model, where every site constitutes a local minimum. In a random trap model shown in **Fig. 1** (B), the transition occurs with equal rate from the site denoted by $i$ to one of the nearest neighbor site $j$ and the rate is given by [12-14]

$$\Gamma_{ij}^{(\text{trap})}(U_i) = \Gamma_0 \exp\left[U_i / (k_B T)\right], \tag{6}$$

where $U_i$ denotes the potential depth and is negative. For the trap model, we consider, in addition to the Gaussian distribution, the case where the site energy distribution is given by an exponential distribution

$$P_{\exp}(U) = \exp\left[-U / (k_B T_0)\right] \tag{7}$$

to elucidate the difference between the results based on the Gaussian distribution and those based on the exponential distribution.

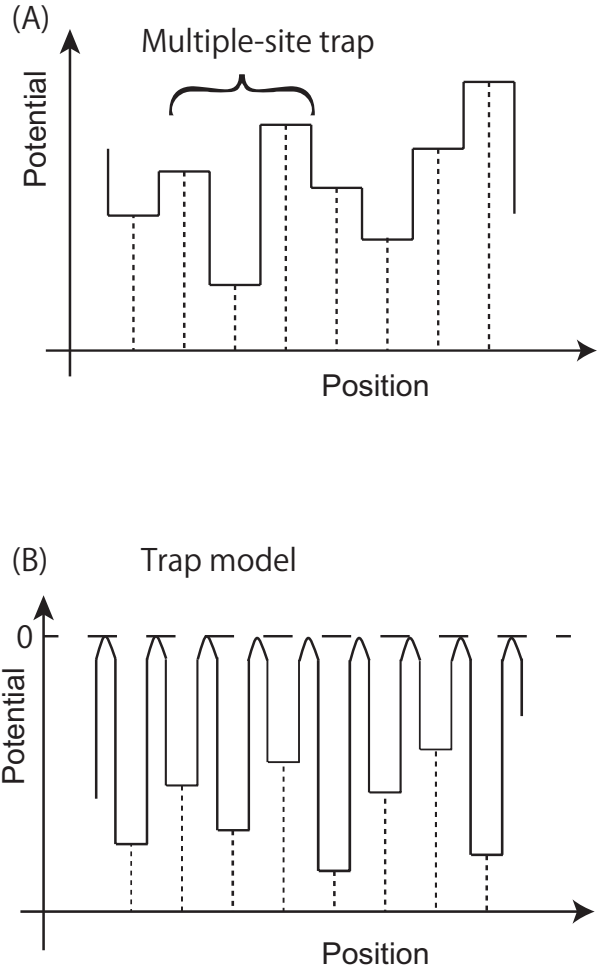

Fig. 1  Schematic representation of (A) rugged energy landscape with Gaussian distribution of energy at sites, showing multiple-site trap model and (B) trap model with random barriers, both in one dimension. The positions of trap sites are denoted by short dashed lines. In (A), site energy can be both maximum and minimum. The 3-site trap can be formed when a local minimum site is surrounded by two local maximum sites in one dimension. In (B), an ordinary trap model is shown, where every trap site constitutes a local minimum. The transition rates are given by the potential depth.

## III.  Treatment of diffusion in one dimensional rugged energy landscape : Quantitative agreement with simulations

Treatment of diffusion in one dimensional systems occupies a special place in studies of random walk mainly because one can obtain exact results for such systems. In one dimension, when a random walker faces a barrier it is forced to bounce back. As a result, when trapped between two large barriers, the random walker traces the same path, repeatedly. The diffusion constant can be expressed formally in terms of all transition rates involved in the trajectory on a line of arbitrary period of $N$ sites. [43] When the random walker is equilibrated after the initial transient period, an exact simple expression of the diffusion constant is obtained using the detailed balance condition, [11-13]

$$D_{\text{eff}} = 1/\langle 1/\left(\rho_i^{(\text{eq})}\Gamma_{ij}\right)\rangle, \qquad (8)$$

where $\rho_i^{(\text{eq})}$ represents the equilibrium distribution at the site denoted by $i$

$$\rho_i^{(\text{eq})} = \frac{\exp\left[-U_i/(k_BT)\right]}{\langle\exp\left[-U_j/(k_BT)\right]\rangle}. \qquad (9)$$

Using Gaussian distribution given by Eq. (3), the effective diffusion constant is obtained as [37]

$$D_{\text{eff}} = D_0 \exp\left[-\sigma^2/\left(k_BT\right)^2\right]/\left[1+\text{erf}\left(\frac{\sigma}{2k_BT}\right)\right], \qquad (10)$$

where $D_0$ is the bare diffusion constant in the absence of rugged landscape. In a very different context, the mobility expression of this type was obtained earlier. [44]

In one dimensional systems, the effective diffusion constant can be generalized to include the spatial correlation given by Gaussian fields characterized by the correlation function

$$\langle U_i U_j \rangle = \sigma^2 \exp\left[-\frac{b^2(j-i)^2}{2\xi^2}\right] , \quad (11)$$

where $\xi$ represents the correlation length and $b$ is the lattice spacing. In the presence of spatial correlation, Eq. (10) is modified. As shown previously, the effective diffusion constant can be obtained using the mean first passage time $\tau_{mft}$ as [11-13]

$$D_{eff} = \lim_{N\to\infty} \frac{1}{D_0} \frac{N^2}{2\tau_{mft}} . \quad (12)$$

Equation (12) reproduces the exact result given by Eq. (8) in the absence of the spatial correlation. In general, the mean first passage time can be expressed as [11-13,15,37]

$$\tau_{mft} = \sum_{i=0}^{N-1} \sum_{j=0}^{i} \left\langle \frac{1}{\Gamma_{i,i+1}} \exp\left[-(U_j - U_i)/(k_B T)\right] \right\rangle . \quad (13)$$

When $U_j$ and $U_i$ are uncorrelated, we can introduce decoupling

$$\tau_{mft} = \sum_{i=0}^{N-1} \sum_{j=0}^{i} \left\langle \frac{1}{\Gamma_{i,i+1}} \exp\left[U_i/(k_B T)\right] \right\rangle \left\langle \exp\left[-U_j/(k_B T)\right] \right\rangle \quad (14)$$

and the effective diffusion constant can be expressed using $\rho_i^{(eq)}$ given by Eq. (9). For correlated Gaussian potential, we cannot introduce decoupling given by Eq. (14) and we need to evaluate Eqs. (12) and (13). The final result is given by [37]

$$D_{eff} = D_0 \exp\left[-\sigma^2/(k_B T)^2\right] \Big/ \left[1 + \mathrm{erf}\left(\frac{\sigma}{2k_B T}\sqrt{1-\exp\left(-\frac{b^2}{2\xi^2}\right)}\right)\right] . \quad (15)$$

By increasing $\xi$, the effective diffusion constant increases and approaches to that of Zwanzig. This is consistent with the fact that the Zwanzig expression of the effective diffusion constant is obtained by introducing an extra smoothing of the rugged landscape. Before closing the section, we stress that the above results are obtained using the special nature of the one dimensional random walk that the path of the random walk can be expressed by all transition rates involved in the random walker's path on a line. An exception is the random trap model, where the effective diffusion constant for the Gaussian potential is given by Eq. (4) in any dimension even under long-range correlations. [14] We will discuss this issue later.

## IV. Effective medium approximation for higher (than one) dimensional systems

Except the trap model, the effective diffusion constant can be calculated only approximately in the dimension higher than one. One of the widely used methods is the effective medium approximation (EMA). [1] In EMA, the random energy landscape is replaced by the effective medium except some sites around the origin. The corresponding Master equation contains a memory kernel with the effective transition rates. The effective transition rate is obtained by imposing that after averaging over the realization of random energy landscape, Green's function of the Master equation should be consistent with that of the Master equation given by the effective transition rate alone.

The simplest EMA uses random site energies connected by a single bond. The random transition rates of the bond should be symmetric for isotropic systems in the absence of external bias. The symmetric transition rates can be constructed via the detailed balance condition using the equilibrium occupation probability. The equilibrium occupation probability at site i denoted by $\rho_i^{(eq)}$ is given by Eq. (9). The symmetric rates in view of the detailed balance can be given by [13]

$$\Gamma_{ij}^{sym} = \rho_i^{(eq)} \Gamma_{ij} \quad . \tag{16}$$

The self-consistency condition in *d*-dimension can be expressed as [12,13,45]

$$\left\langle \frac{\Gamma_{eff} - \Gamma^{sym}}{(d-1)\Gamma_{eff} + \Gamma^{sym}} \right\rangle = 0, \tag{17}$$

where $\Gamma_{eff}$ denotes the effective mobility.

In one dimension, the result can be simplified as [13]

$$\frac{1}{\Gamma_{eff}} = \left\langle \frac{1}{\Gamma^{sym}} \right\rangle, \tag{18}$$

which leads to the exact result given by Eq. (8). In general, the analytical exact solution of the self-consistency condition is not available in the dimension higher than one. We solved the self-consistency equation approximately and compared the solution with the numerical results of the self-consistent equation.

In order to solve the self-consistency condition approximately, we note that the self-consistency condition can be rewritten as, [38]

$$\frac{1}{d} = \left\langle \frac{1}{1+(d-1)\Gamma_{\text{eff}}/\Gamma^{\text{sym}}} \right\rangle . \tag{19}$$

The right-hand side of eq. (19) can be smaller as $\Gamma^{\text{sym}}$ is smaller, which is suited for the application of the saddle point method. By defining

$$f(x,y) = \frac{1}{2\pi\sigma^2} \exp\left(-\frac{x^2+y^2}{2\sigma^2}\right) g(y) , \tag{20}$$

where $g(y)$ is defined by

$$g(y) = \left[1 + \frac{(d-1)\Gamma_{\text{eff}}}{\Gamma_0} \exp\left(\frac{\sigma^2}{2(k_BT)^2} + \frac{y}{k_BT}\right)\right]^{-1} , \tag{21}$$

eq. (19) can be rewritten using $\Delta U_i = U_j - U_i$ as

$$\begin{aligned}\frac{1}{d} &= \int_{-\infty}^{0} d\Delta U_i \int_{-\infty}^{\infty} dU_i f(U_i+\Delta U_i, U_i) + \int_{0}^{\infty} d\Delta U_i \int_{-\infty}^{\infty} dU_i f(U_i, U_i+\Delta U_i) \\ &= \int_{0}^{\infty} d\Delta U_i \int_{-\infty}^{\infty} dU_i \left[f(U_i-\Delta U_i, U_i) + f(U_i, U_i+\Delta U_i)\right] \\ &= 2\int_{0}^{\infty} d\Delta U_i \int_{-\infty}^{\infty} dU_i f(U_i-\Delta U_i, U_i),\end{aligned} \tag{22}$$

where the last equality follows from $f(U_i, U_i+\Delta U_i) = f(y-\Delta U_i, y)$ with $y = U_i + \Delta U_i$ and by changing the integration variable from $U_i$ to $y$. By using

$$\int_0^{\infty} d\Delta U_i \frac{1}{\sqrt{2\pi\sigma^2}} \exp\left(-\frac{(U_i-\Delta U_i)^2}{2\sigma^2}\right) = \frac{1}{2}\left[1+\text{erf}\left(\frac{U_i}{\sqrt{2\sigma^2}}\right)\right], \tag{23}$$

we obtain,

$$\frac{1}{d} = \int_{-\infty}^{\infty} dU_i \frac{1}{\sqrt{2\pi\sigma^2}} \exp\left(-\frac{U_i^2}{2\sigma^2}\right) \frac{1+\text{erf}\left(U_i/\sqrt{2\sigma^2}\right)}{1+\exp\left[(U_i-\mu)/(k_BT)\right]} , \tag{24}$$

where $\mu$ is defined by,

$$\mu = -k_B T \ln\left[(d-1)(\Gamma_{eff}/\Gamma_0)\right] - \frac{\sigma^2}{2k_B T}. \tag{25}$$

Since the factor $1/\{1+\exp[(U_i-\mu)/(k_B T)]\}$ is close to $1$ when $U_i$ is up to $\mu$ and decreases to zero as $U_i$ increases over $\mu$, Eq. (24) can be approximated as,

$$\frac{1}{d} = \int_{-\infty}^{\mu} dU_i \frac{1}{\sqrt{2\pi\sigma^2}} \exp\left(-\frac{U_i^2}{2\sigma^2}\right)\left[1+\mathrm{erf}\left(U_i/\sqrt{2\sigma^2}\right)\right]$$
$$= \frac{1}{4}\left[1+\mathrm{erf}\left(\frac{\mu}{\sqrt{2\sigma^2}}\right)\right]^2. \tag{26}$$

We note that $\mu$ defined by Eq. (25) is close to zero by introducing $\Gamma_{eff}$ obtained numerically using the original self-consistent condition. By introducing the approximation given by $\mathrm{erf}(x) \approx 2x/\sqrt{\pi}$ when $x \sim 0$, we obtain,

$$\frac{1}{d} = \frac{1}{4}\left[1+\frac{\sqrt{2}}{\sqrt{\pi\sigma^2}}\mu\right]^2. \tag{27}$$

By substituting eq. (25) into eq. (27), and rearrangement we obtain,

$$\frac{\Gamma_{eff}}{\Gamma_0} = \frac{1}{d-1}\exp\left[-\sigma^2/(2k_B^2 T^2) - \frac{\sqrt{2\pi\sigma^2}}{k_B T}\left(\frac{1}{\sqrt{d}}-\frac{1}{2}\right)\right]. \tag{28}$$

In 2 d, the result can be expressed as,

$$\frac{\Gamma_{eff}}{\Gamma_0} = \exp\left[-\frac{\sigma^2}{2(k_B T)^2} - \left(1-\frac{1}{\sqrt{2}}\right)\frac{\sqrt{\pi}\sigma}{k_B T}\right]. \tag{29}$$

According to the Einstein relation, the ratio between the effective diffusion constant and the bare diffusion constant is given by the ratio between the effective transition rate and the bare transition rate $D_{eff}/D_0 = \Gamma_{eff}/\Gamma_0$.

## V. Results and analysis

In **Fig. 2**, we present the results of theoretical analyses, along with limited amount of available simulation results. In Fig. 2, 1 dimensional exact result is obtained from Eq. (10). We also show Zwanzig's results by red dashed line. The open circles and squares are obtained by numerical solution of the self-consistency equation Eq. (17). Thick black dashed line indicates the 2-dimensional approximate result of Eq. (29). The dashed-and-dotted line represents Eq. (28). In 3D the result of Eq. (28) did not reproduces the correct limit $\Gamma_{eff}/\Gamma_0 \to 1$ as $\sigma^2 \to 0$. The error occurred in deriving Eq. (26) when we introduced $\mu$ and the integration to $\infty$ is set up to $\mu$ and assumed $\mu \sim 0$. By slightly modifying Eq. (28) derived using EMA in 2 d and 3 d, analytical results can be well approximated by

$$\frac{\Gamma_{eff}}{\Gamma_0} = \exp\left[-\frac{\sigma^2}{2(k_B T)^2} - \left(\sqrt{\frac{2}{d}} - \frac{1}{\sqrt{2}}\right)\frac{\sqrt{\pi}\sigma}{k_B T}\right]. \tag{30}$$

In 2 d, the result is the same as that of Eq. (29). In 3 d, the modified result is shown in the magenta thick line. The results are close to published Monte-Carlo simulation results.

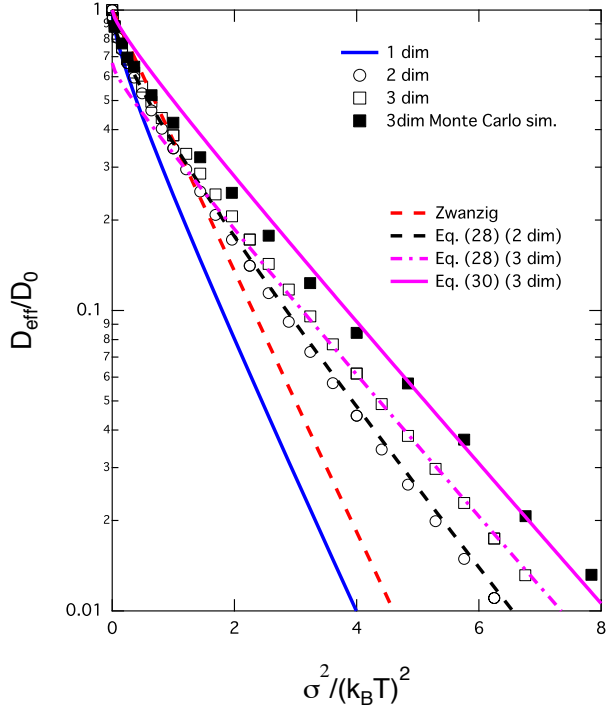

**Fig. 2** $D_{eff}/D_0$ is plotted as a function of $\sigma^2/(k_B T)^2$. The numerical solutions of the self-consistency equation given by Eq. (17) are shown by circles (2D), and squares (3D). The thick blue line indicates the analytical exact result of 1D given by Eq. (10). The approximate results of Eq. (28) are given by the thick dashed line (2D) and the magenta dash-dotted line (3D). The magenta thick line represents the results of Eq. (30) for 3D. The red thick dashed line indicates Zwanzig's expression given by Eq. (4). The black squares represent the Monte-Carlo simulation results of ref. [13]. In one dimension, the theoretical result agrees quantitatively with simulations [38]. Zwanzig's expression moderately overestimates the exact result in 1D.

**Figure 2** clearly brings out the different nature of diffusion in one dimension compared to that in higher dimensional systems. For example, when $\sigma^2/(k_B T)^2 = 6$, $D_{eff}/D_0$ in one dimension is order of magnitude smaller than those in two and three dimensions.

# VI. The upper bound of the effective diffusion constant

In high dimensions, the transitions are most likely considered to be independent events. Even when the transition rate to a certain neighboring site is extremely small, the transition back to the previously occupied site is unlikely in 2 d or higher dimensions. In the simplest effective medium approximation, correlations between two sites are taken into account. This implies that the random walker remembers the transition rate at the previous step on the paths. In the limit of $d \to \infty$, the memory of previous jumps is lost. The loss of memory about the previous step by taking $d \to \infty$ limit is taken into account in the EMA result. Indeed, by taking $d \to \infty$ limit of Eq. (17), we obtain the exact result

$$\Gamma_{\text{eff}} \approx \langle \Gamma^{\text{sym}} \rangle = \Gamma_0 \text{erfc}\left(\frac{\sigma}{2k_B T}\right) \quad (31)$$

By using the asymptotic expansion,

$$\text{erfc}(x) \approx \frac{\exp(-x^2)}{\sqrt{\pi} x}, \quad (32)$$

the results can be approximated as,

$$\frac{\Gamma_{\text{eff}}}{\Gamma_0} \approx \frac{2k_B T}{\sqrt{\pi}\sigma} \exp\left(-\frac{\sigma^2}{4(k_B T)^2}\right) \quad (33)$$

when $\sigma/(2k_B T) \gg 1$. Equation (31) is derived when the each transition is statistically independent. The situation is in sharp contrast to the transition in 1d. In 1d, if transition to a new position does not occur, the position after the transition should be the same place previously occupied. The each transition is not independent and the result given by eq. (10) differs from eq. (31). In 2 and 3 d, such correlation is partly preserved and the resultant equation given by eq. (30) also differs from eq. (31).

In the below, we show that the value of $\Gamma_{eff}/\Gamma_0$ given by eq. (31) is the upper bound.

First we rewrite eq. (17) as

$$d\Gamma_{eff} = 1 \bigg/ \left\langle \frac{1}{(d-1)\Gamma_{eff} + \Gamma^{sym}} \right\rangle. \tag{34}$$

Using Jensen's inequality given by

$$\int_0^\infty ds \exp\left[-s\left\langle \frac{1}{(d-1)\Gamma_{eff} + \Gamma^{sym}} \right\rangle\right] \leq \left\langle \int_0^\infty ds \exp\left[-s\frac{1}{(d-1)\Gamma_{eff} + \Gamma^{sym}}\right] \right\rangle, \tag{35}$$

we obtain

$$1 \bigg/ \left\langle \frac{1}{(d-1)\Gamma_{eff} + \Gamma^{sym}} \right\rangle \leq (d-1)\Gamma_{eff} + \left\langle \Gamma^{sym} \right\rangle. \tag{36}$$

By combining eqs. (34) and (36), we have

$$d\Gamma_{eff} \leq (d-1)\Gamma_{eff} + \left\langle \Gamma^{sym} \right\rangle. \tag{37}$$

By rearrangement we find

$$\Gamma_{eff} \leq \left\langle \Gamma^{sym} \right\rangle. \tag{38}$$

The result indicates that $\langle \Gamma^{sym} \rangle$ is the upper bound of $\Gamma_{eff}$ calculated from the self-consistency equation. If a random walker is surrounded by high barriers and barriers are static, the diffusion is suppressed compared to the situation that the barriers are not static and can be removed. If every transition is treated as statistically independent event, barrier height is determined for each transition and is a dynamical quantity. As a result, the effective transition rate overestimates that of a random walker under quenched disorder.

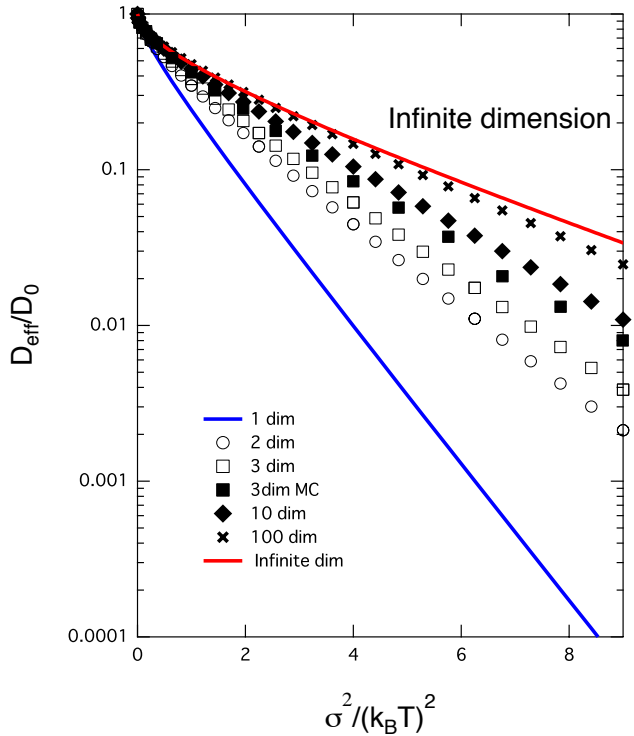

**Fig. 3** $D_{eff}/D_0$ **as a function of** $\sigma^2/(k_B T)^2$. **The thick red line indicates the result of infinite dimension given by Eq. (31). The numerical solutions of the self-consistency equation given by Eq. (17) are shown by circles (2d), squares (3d), diamonds (10d) and crosses (100d). The thick blue line indicates the analytical exact result of 1d given by Eq. (10). The black squares represent the Monte-Carlo simulation results of ref. [13].**

In **Figure 3**, we show the results of infinite dimension given by Eq. (31). The results are the upper bounds of $D_{eff}/D_0$ as we have proven above. When $\sigma^2/(k_B T)^2 = 6$, $D_{eff}/D_0$ of infinite dimension is in the same order as those of 2 and 3 d. In 1 d, if the transition rate to the new site is small due to the energy barrier, random walker moves back to the original occupied site. In infinite dimension, even if the transition to a certain site requires high energy barrier, there would always be another available site with a lower energy barrier. These are two opposite limits. As shown in Fig. 3, $D_{eff}/D_0$ in one dimension differs significantly from those in higher dimensions. However, even in high dimensions, effects of ruggedness remains significant.

## VII. Random trap model

In the random trap model, the transition rate to a neighboring site is the same for any direction in any dimension. Even for the dimension higher than one, the effective diffusion constant is exactly given by Zwanzig's effective diffusion constant. [14] This can be understood from the fact that symmetrized rate is given by the bare transition rate divided by $\langle \exp[-U_i/(k_B T)] \rangle$ at every site, [14]

$$\Gamma_{ij}^{sym} = \rho_i^{(eq)} \Gamma_{ij}^{trap} = \Gamma_0 / \langle \exp[-U_i/(k_B T)] \rangle. \tag{39}$$

By substituting $\Gamma_{ij}$ calculated from Eq. (39) into the one dimensional result of Eq. (8),

Zwanzig's expression of Eq. (4) can be obtained. Since $\Gamma_{ij}^{sym}$ is a constant, the effective diffusion constant is given by Zwanzig's expression even in the dimension higher than 1d. The above relation on the symmetrized rate holds even under long-range correlations. It can be easily shown that the effective diffusion constant is still given by Zwanzig's result of Eq. (4) under a long-range correlation.

For the trap model, if the average in the above equation [Eq. (8)] is evaluated using exponential ruggedness given by Eq. (7), the diffusion constant becomes zero when $\alpha = T/T_0$ is smaller than one. In this case, the time evolution of the mean square displacement is given by

$$\langle r^2(t) \rangle = 2dD_{eff}^{(\gamma)} t^\gamma \tag{40}$$

with the exponent $\gamma$ being smaller than one. [46] The process is called sub-diffusion. If the random walk is sub-diffusive, the trajectories are localized compared to those of normal diffusion at long times. For this case, the exact solution of the form Eq. (8) vanishes. However, *we obtain $\gamma = 2\alpha/(\alpha+1)$ in one dimension and $\gamma = \alpha$ in dimension higher than 1* apart from a weak logarithmic correction term in 2d, when the density of states is expressed by the exponential distribution [Eq. (7)]. [46] In 1d, $\gamma = 2\alpha/(\alpha+1)$ can also be obtained by using the random barrier model. [8,47] In the random barrier model, the site energies are the same and the activation energies for the transitions are random. In one dimension, the self-consistency equation expressed in the Laplace domain is the same as that obtained by the random trap model including the dependence on the Laplace variable. Indeed, the duality between the random barrier model and the random trap model in one dimension has been rigorously proved. [36] The mean square displacements of the dual models are equal on all time scales. Therefore, in the following, we do not distinguish between the results of the random trap model and those of the random barrier model in 1d.

In 1d, it is also known that transient kinetics obtained by the EMA can be expressed using hypernetted chain (HNC) diagram. [48] In 1d, EMA value of $D_{eff}^{(\gamma)}$ differs from that obtained using more accurate calculations. [49,50] In Fig. 4(a), the solid line indicates the EMA result given by [12]

$$D_{eff}^{(\gamma)} / \left( b^2 \Gamma_0^\gamma \right) = \left[ \sin \pi\alpha / \left( 2^{1-\alpha} \pi\alpha \right) \right]^{2/(1+\alpha)} / \Gamma\left[ (1+3\alpha)/(1+\alpha) \right]. \qquad (41)$$

The deviation of the EMA results of $D_{eff}^{(\gamma)}$ from the more accurate results can be seen in Fig. 4(a) when the reduced temperature $\alpha = T/T_0$ is decreased from 0.5. Although the value of $D_{eff}^{(\gamma)}$ deviates by decreasing the value of $\alpha$, the diagonal element of the Green function obtained by EMA works fairly accurately for any value of $\alpha$. [50] In Fig. 4 (b), the solid line represents the prefactor of the asymptotic time dependence for the diagonal element of the Green function [50]

$$C_{GF} = 1/\lim_{t\to\infty}\left(G_{00}(t)\Gamma(1-\gamma/2)t^{\gamma/2}\right), \tag{42}$$

where the diagonal element of the Green function for the random trap model in 1d is denoted by $G_{00}(t)$. In EMA, the probability density profile of random walkers is expressed by a Gaussian. The deviation of $D_{eff}^{(\gamma)}$ indicates that the profile deviates from this Gaussian by decreasing the value of $\alpha$. Judging from Fig. 4 (a) and (b), the profile of random walkers can be deviate from a Gaussian when $\alpha$ is smaller than 0.5.

For the higher dimension [d>1], the duality does not hold and the mean square displacements can be calculated only approximately. For random trap models, EMA for site disorder should be used instead that for bond disorder considered so far. In d dimensions and for lattice disorder at the origin, the self-consistent equation of EMA with the Laplace transform of the time variable can be expressed as [12]

$$\langle \frac{\Gamma_{eff}(s) - \Gamma^{trap}}{\Gamma^{trap} + sE_0(s)\left[\Gamma_{eff}(s) - \Gamma^{trap}\right]}\rangle = 0 \quad, \tag{43}$$

where $\Gamma_{eff}(s)$ is the effective transition rate expressed in the Laplace domain and $\Gamma^{trap}$ is the transition rate at the origin to an arbitrary neighboring lattice site. For random trap models, $\Gamma^{trap}$ depends only on the lattice energy at the origin and is independent of the energy at the destination site of the transition. In the above, we define

$$E_0(s) = g\left(0, \frac{2d\Gamma_{eff}(s)}{s + 2d\Gamma_{eff}(s)}\right) \Big/ (s + 2d\Gamma_{eff}(s)), \tag{44}$$

where the generating function of the lattice Green function is given by

$$g(\mathbf{r}, \xi) = (2\pi)^{-d} \int_B d\mathbf{k} \exp(-i\mathbf{r} \cdot \mathbf{k}) / \left[1 - \xi\lambda_{str}(\mathbf{k})\right]. \tag{45}$$

$B$ represents the first Brillouin zone and $\lambda_{str}$ is the lattice structure factor. For a simple cubic lattice we have $g(0,1) = 1.516386$. [12]

As shown in the Appendix, the effective diffusion coefficients can be obtained as

$$D_{eff}^{(\gamma)} / \left(b^2 \Gamma_0^\gamma\right) = \frac{\sin \pi\alpha}{\pi\alpha} \left(\frac{\ln(32)}{4\pi}\right)^{1-\alpha} \frac{t^\alpha}{\Gamma(1+\alpha)}, \quad \text{(2d)} \tag{46}$$

$$D_{eff}^{(\gamma)} / \left(b^2 \Gamma_0^\gamma\right) = \frac{\sin \pi\alpha}{\pi\alpha} \left(\frac{g(0,1)}{6}\right)^{1-\alpha} \frac{t^\alpha}{\Gamma(1+\alpha)}, \quad \text{(3d)} \tag{47}$$

for 2d (apart from a logarithmic factor) and 3d, respectively, where $\gamma = \alpha$ in both dimensions.

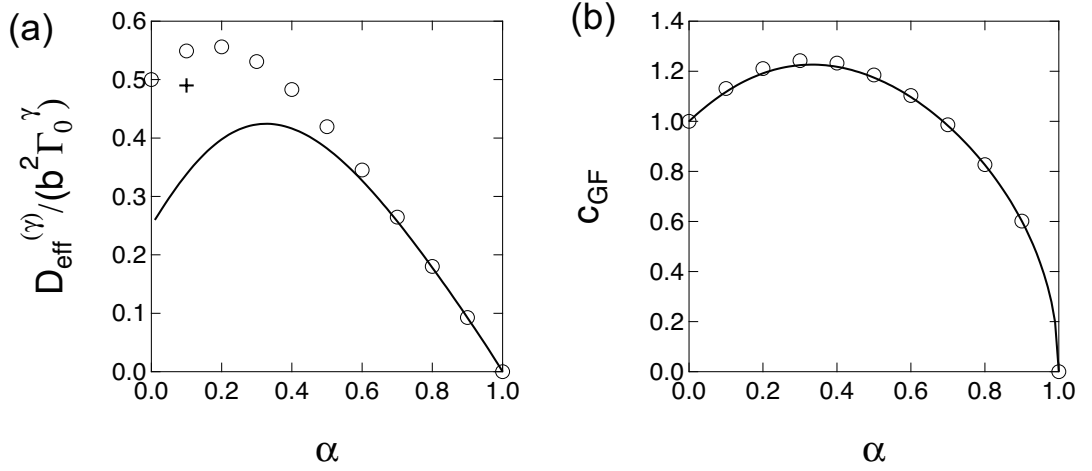

**Fig. 4** (a) $D_{eff}^{(\gamma)}/(b^2\Gamma_0^\gamma)$ as a function of $\alpha$. (b) $C_{GF} = 1/\lim_{t\to\infty}\left(G_{00}(t)\Gamma(1-\gamma/2)t^{\gamma/2}\right)$ as a function of $\alpha$. $\gamma = 2\alpha/(\alpha+1)$. The effective diffusion constant and the diagonal element of the Green function for the random trap model in 1d are expressed by $D_{eff}^{(\gamma)}$ and $G_{00}(t)$, respectively. The solid lines indicate the EMA results given by Eq. (41) and Eq. (42), respectively. The circles represent the accurate result of ref. [50]. In (a), the results of ref. [49] overlap those given by circles. The cross in (a) indicates the numerical result of ref. [51].

If the transition to one of the nearest neighbor sites is considered as statistically independent, the waiting time distribution of making a jump to a nearest neighbor site can be expressed as

$$\psi(t) = \int_0^\infty dU P_{exp}(U)\Gamma_{ij}^{(trap)}(U)\exp\left[-2d\Gamma_{ij}^{(trap)}(U_i)t\right] \sim 1/t^\alpha. \qquad (48)$$

Continuous time random walk (CTRW) can be formulated using the waiting time distribution and we obtain $\gamma = \alpha$. [52] As far as the exponent $\gamma$ is concerned, CTRW formalism is a reasonable approximation when the dimension is higher than one.

The exponent $\gamma$ of the sub-diffusion processes of the one dimensional random walk is different from that in the dimension higher than one. The difference indicates that the random walk trajectories in one dimension differ from those in other dimensions. In one dimension, the random walker moves back and forth and the movement produces larger temporal correlations.

## VIII. Conclusion

One is well-aware that in isotropic homogeneous medium, diffusion constant is usually independent of dimension since the dimension $d$ is factored out by dividing the mean square displacement by $d$, in Einstein's expression. This may be the reason why (as discussed extensively by Stein and Newman) most of the existing discussions of relaxation in random systems employ a description of one dimensional diffusion.[39-41] The underlying physical picture is that of a random walker in one dimensional random, dissipative environment where the relaxation behavior is insensitive to dimension.

Thus, one often "freely" uses reduction in degrees of freedom, as common in time dependent statistical mechanics, to obtain a one dimensional description, such as the one employed by van der Zwan and Hynes in the study of various chemical reactions.[53] Such a reduction procedure leaves one with a frequency dependent friction in the Langevin equation which can then be transformed into a time dependent diffusion equation, if needed. [54] Another classic example of such reduction procedure is Zwanzig's treatment of coupled oscillators where the solute degree of freedom is coupled to N-number of harmonic oscillators.[55]  As in the previous examples, this reduction of degrees of freedom is accommodated through a frequency dependent friction term.

The present analysis suggests that the "effective" landscape that a protein may experience on its journey towards the native state is smoothened by higher dimensionality. Thus, multidimensionality is a way to understand the "Principle of minimum frustration" advocated by Wolynes, Onuchic and co-workers [25,56-59]. As a walker can get trapped in 1d, having an extra dimension really can help in making the effective landscape (that is the landscape explored) smooth and minimally frustrated.

As already noted, Stein-Newman's argument about pathological nature of diffusion in a one dimensional rugged landscape (with simultaneous presence of maxima and minima that enhances the chance of trapping of the random walker) arises from the observation that the chance of encountering larger and larger barriers increase as $T \log t$ where $T$ is the temperature and $t$ is the time. The random walker faced with a large barrier gets rebounded and then retraces the same path it traversed before, leading to lowering and eventual vanishing of diffusion constant. This can be understood from Eq.1.

In this context, another result of Stein and Newman that is of particular relevance here is that when dimensionality is higher than eight, different paths "from the lake leading to the sea" need not overlap. This conclusion is reached by mapping the random walk in random environment (RWRE) problem to an invasion percolation problem. [41,43] Note that conclusions of N-S are strictly valid in the limit when the T going to zero limit is taken prior to t going to infinity limit. But nevertheless, their conclusions have relevance when ruggedness is large and rate limiting, as discussed previously and demonstrated in this work.

The result that diffusion in rugged landscape can be drastically different in 1d from that in higher dimensions is in itself an interesting result, so is the non-trivial dependence of diffusion constant on the dimension d. *This is purely a consequence of ruggedness*. Such a scenario also unfolds in random trap model with an exponential distribution of activation energies in the escape rates. However, in the latter case we do not recover diffusion even in the long time limit. In the present case of Gaussian ruggedness, no such difficulty arises and diffusion constant exists if we keep ruggedness fixed and take the limit t going to infinity properly.

*Our result clearly demonstrates that existence of ruggedness makes diffusion of a particle strongly dimensionality dependent.* In particular, diffusion increases by even a factor of 5 at intermediate level of ruggedness in going from 1d to 2d. This seems to vindicate the argument of Stein and Newman about the unusual constrain that a random walker faces in 1d.

Compared to the Gaussian distribution of energies, diffusion in 1d is even more different from that in higher dimension for a heavy-tailed ruggedness given by an exponential energy distribution in random trap models. In quenched disorder with an exponential distribution, the mean square displacements grow sublinearly with time when the temperature is below a certain threshold value. [8,12] The exponent of the sublinear growth in 1d differs from those in higher dimensions. [8,46] The strong dimensional dependence is induced by correlations in random walk trajectories in 1d as explained above. If the random walker faces with a large barrier, the walker tends to retrace the same path traveled before in 1d. The correlations increase by decreasing the dimensionality and the critical dimension is two as regards to the growth exponents of the mean square displacements. [8,46]

In the results presented here and elsewhere [37], we show that presence of positive correlations (giving coherence among energy values) *increase* diffusion in one dimensional systems. The magnitude of the effect is small in the model studied which may be a consequence of 1d and/or simplicity of the model.

It may be interesting to explore this role of correlations in more details. This has been a topic of discussion in evolutionary biology [33]. It will be particularly interesting to explore the effect of anti-correlation among energy landscape that may decrease the value of diffusion significantly. Correlations in landscape may explain the crossover from Rosenfeld to Adam-Gibbs scenario. We are working on this problem.

It remains an interesting unfinished work to obtain a description of the time dependence of diffusion, or time dependence of mean square displacement, of the random walker in an arbitrary dimension, as a function of ruggedness (or, temperature, T) and time, t. In the short-to-intermediate times (to be determined by ruggedness in each case), dynamics lacks universality and determined by specificity of the model.

As future problems, more extensive computer simulation studies of dimensionality dependence of diffusion in correlated landscapes shall be worthwhile pursuits. We particularly need simulation results on diffusion in rugged two dimensional energy landscapes. No simulation results seem to exist for this system. Role of spatial correlations certainly deserves further work, not only just the length scale dependence but also the nature. Lastly, the role of temperature on diffusion in rugged landscape requires special attention, not just in the context of the discussed cross-over from Rosenfeld scaling regime to Adam-Gibbs scenario but also from a fundamental dynamical point of view such as broken or compromised ergodicity in this model. Emergence of spatial correlations in the energy distribution of the landscape may play an important role in determining this crossover.

**ACKNOWLEDGEMENT**

We thank Professors Peter Wolynes and Jose Onuchic for interesting discussions. K. Seki acknowledges support from JSPS KAKENHI Grant Number 15K05406. B. Bagchi acknowledges support from Department of Science and Technology (DST) and also from Sir J.C. Bose fellowship, India. BB also thanks Prof. Iwao Ohmine and Prof. Shinji Saito for hospitality at IMS, Okazaki, Japan.

**APPENDIX. EMA FOR RANDOM TRAP MODELS**

For convenience, we define,

$$\Gamma_{eff}^{(r)}(s) = \frac{\Gamma_{eff}(s)}{\Gamma_0} \quad . \tag{A1}$$

By evaluating the ensemble average using exponential form of the site energy given by Eq. (7), the self-consistent equation can be rewritten as,

$$\Gamma_{eff}^{(r)}(s) = \frac{\alpha}{1+\alpha} \frac{{}_2F_1\left[1,1+\alpha;2+\alpha;-(1-sE_0(s))/\left(s\Gamma_{eff}^{(r)}(s)E_0(s)\right)\right]}{{}_2F_1\left[1,\alpha;1+\alpha;-(1-sE_0(s))/\left(s\Gamma_{eff}^{(r)}(s)E_0(s)\right)\right]} \quad , \tag{A2}$$

where ${}_2F_1(a,b;c;x)$ represents the Gaussian hypergeometric function [60]. When $\alpha < 1$ and $s \to 0$, we can applying asymptotic expansion [60]

$$ {}_2F_1[1,\alpha;1+\alpha;-x] \approx \frac{\pi\alpha}{\sin\pi\alpha} x^{-\alpha} \quad , \tag{A3}$$

$$ {}_2F_1[1,1+\alpha;2+\alpha;-x] \approx \frac{\alpha+1}{\alpha}\frac{1}{x} \quad , \tag{A4}$$

for $x \to \infty$. In these cases, Eq. (A2) can be approximately expressed as

$$\left(\frac{\Gamma_{eff}(s)}{\Gamma_0}\right)^\alpha \approx \frac{\sin\pi\alpha}{\pi\alpha}(sE_0(s))^{1-\alpha} \quad . \tag{A5}$$

The dimensionality dependence of $\Gamma_{eff}(s)$ originates from the difference in the generating function of the lattice Green function in $E_0(s)$ defined by Eq. (44).

In 1d, we have $g(0,\xi) = 1/\sqrt{1-\xi^2}$ [12] and

$$sE_0(s) \approx \frac{1}{2}\sqrt{\frac{2}{\Gamma_{eff}(s)}} \tag{A6}$$

when $s \to 0$. By substituting the above result into Eq. (A5), we obtain,

$$\Gamma_{eff}(s) \approx \Gamma_0^{2\alpha/(1+\alpha)} \left(\frac{\sin\pi\alpha}{2^{1-\alpha}\pi\alpha}\right)^{2/(1+\alpha)} s^{(1-\alpha)/(1+\alpha)} \quad . \tag{A7}$$

The mean square displacements are obtained from the inverse Laplace transform of

$2db^2\Gamma_{eff}(s)/s$ as

$$\langle r^2(t)\rangle = 2b^2\left(\frac{\sin\pi\alpha}{2\pi\alpha}\right)^{2/(1+\alpha)}(2\Gamma_0 t)^{2\alpha/(1+\alpha)}/\Gamma\left(\frac{1+3\alpha}{1+\alpha}\right). \tag{A8}$$

The result leads to Eq. (41).

In 2d, we have $g(0,\xi)=(1/\pi)\ln[8/(1-\xi)]$ [12] and

$$sE_0(s) \approx \frac{s}{4\pi\Gamma_{eff}(s)}\ln\left[\frac{32\Gamma_{eff}(s)}{s}\right] \tag{A9}$$

when $s\to 0$. By substituting the above result into Eq. (A5), we obtain,

$$\frac{\Gamma_{eff}(s)}{s} \approx \Gamma_0^\alpha \frac{\sin\pi\alpha}{\pi\alpha}\frac{1}{(4\pi)^{1-\alpha}s^\alpha}\left[\ln\left(\frac{32\Gamma_{eff}(s)}{s}\right)\right]^{1-\alpha}. \tag{A10}$$

The above equation is a self-consistent equation for $\Gamma_{eff}(s)/s$. Apart from a weak logarithmic factor, the mean square displacements are obtained from the inverse Laplace transform of $2db^2\Gamma_{eff}(s)/s^2$ as

$$\langle r^2(t)\rangle = 4b^2\frac{\sin\pi\alpha}{\pi\alpha}\left(\frac{\ln(32)}{4\pi}\right)^{1-\alpha}\frac{(\Gamma_0 t)^\alpha}{\Gamma(1+\alpha)}. \tag{A11}$$

By comparing the result with Eq. (40), we obtain Eq. (46).

In 3d, we can express

$$sE_0(s) \approx \frac{s}{6\Gamma_{eff}(s)}g(0,1), \tag{A12}$$

when $s\to 0$. By substituting Eq. (A12), Eq. (A5) can be rewritten as,

$$\frac{\Gamma_{eff}(s)}{\Gamma_0^\alpha} \approx \frac{\sin\pi\alpha}{\pi\alpha}\left(\frac{s}{6}g(0,1)\right)^{1-\alpha}. \quad (A13)$$

The mean square displacements can be obtained from the inverse Laplace transform of $2db^2\Gamma_{eff}(s)/s^2$. Using Tauberian theorem, we obtain

$$\langle r^2(t)\rangle = 6b^2 \frac{\sin\pi\alpha}{\pi\alpha}\left(\frac{g(0,1)}{6}\right)^{1-\alpha}\frac{(\Gamma_0 t)^\alpha}{\Gamma(1+\alpha)}. \quad (A14)$$

By comparing the result with Eq. (40), we obtain Eq. (47).

As shown in the main text, $\langle r^2(t)\rangle \sim t^{2\alpha/(1+\alpha)}$ is obtained for 1d and $\langle r^2(t)\rangle \sim t^\alpha$ is obtained for 2d and 3d. As regards to the exponents, they were obtained by different methods [46].

**REFERENCES**

1. J.M. Ziman, "Models of Disorder" (Cambridge, Cambridge/London, 1979).

2. B. Bagchi, "Molecular Relaxation in Liquids" (Oxford, New York, 2012).

3. D. J. Wales, Energy Landscapes (Cambridge university press, Cambridge, 2003).

4. H. Scher and M. Lax, Phys. Rev. B **7**, 4491 (1973).

5. H. Scher and M. Lax, J. Non-Crystal. Solids, **8/10**, 497 (1972).

6. H. Scher and E. W. Montroll, Phys. Rev. B **12**, 2455 (1975).

7. E. W. Montroll and G. H. Weiss, J. Math. Phys. **6**, 167 (1965).

8.  J. Bernasconi, H. U. Beyeler, S. Strässler, S. Alexander, Phys. Rev. Lett. **42**, 819 (1979).

9.  A. Miller and E. Abraham, Phys. Rev. **120**, 745 (1960).

10. V. Ambegaokar, B. I. Halperin, and J. S. Langer, Phys. Rev. B **4**, 2612 (1971).

11. K.P.N. Murthy, and K.W. Kehr, Phys. Rev. A **40**, 2082 (1989); Erratum ibid.**41**, 1160 (1990).

12. J. W. Haus and K. W. Kehr, Phys. Rep. **150**, 263 (1987).

13. K.W. Kehr, T. Wichmann, "Diffusion Coefficients of Single and Many Particles in Lattices with Different Forms of Disorder", Materials Science Forum, **223-224**, 151 (1996).

14. Haus, J. W., Kehr, K. W. & Lyklema, J. W. Phys. Rev. B **25**, 2905 (1982).

15. K. Seki and M. Tachiya, Phys. Rev. B **65**, 014305 (2001).

16. G.P. Johari and M. Goldstein, J. Chem. Phys. **53**, 2372 (1970).

17. S. Sastry, P. G. Debenedetti and F. H. Stillinger, Nature **393**, 554 (1998).

18. S. Sastry, Nature **409**, 164 (2001).

19. F.H. Stillinger and T.A.Weber, J. Chem.Phys. **81**, 5095 (1984).

20. T. Keyes, J. Phys. Chem. A **101**, 2921 (1997).

21. W-X. Li and T. Keyes, J. Chem. Phys. **111**, 5503 (1999).

22. V. K. de Souza and D. J. Wales, J. Chem. Phys. **129**, 164507 (2008).

23. V. K. de Souza and D. J. Wales, J. Chem. Phys. **130**, 194508 (2009).

*24.* J. N. Onuchic, Z. Luthey-Schulten and P.G. Wolynes, Annu. Rev. Phys. Chem. 18, 545 (1997), (*See Figure 3 for an illustration of random energy model)*

25. (a) J. D. Bryngelson and P. G. Wolynes, J. Phys. Chem. **93**, 6902 (1989).; (b) J. Wang, S. S. Plotkin, and P. G. Wolynes, J. Phys. I France **7**, 395 (1997).

26. R. Ghosh, S. Roy and B. Bagchi, J. Chem. Phys. **141**, 135101 (2014).

27. R. Zwanzig, Proc. Natl. Acad. Sci. **85**, 2029 (1988).

28. S. Lifson and J.L. Jackson, J. Chem. Phys. **36**, 2410 (1962).

29. M. Slutsky and L. A. Mirny, Biophys J. **87**, 4021 (2004).

30. B. Bagchi, P. C. Blainey, and X. S. Xie, *J. Phys. Chem. B* **112**, 6282 (2008).

31. P. C. Blainey, G. Luo, S. C. Kou, W. F. Mangel, G. L. Verdine, B. Bagchi, and X. S. Xie, Nat Struct Mol Biol **16**, 1224 (2009).

32. (a) W Min, X. S. Xie and B. Bagchi, J. Phys. Chem. B **112**, 454    (2008). (b) W.Min, X. S. Xie, and B. Bagchi, *J Chem Phys* **131**, 065104.A (2009).

33. E. Weinberger, Biol. Cybermatics, 63, 325-336 (1990).

34. (a) Y. Rosenfeld, Chem. Phys. Lett. **48**, 467 (1977). (b) Y. Rosenfeld, Phys. Rev. A **15**, 2545 (1977).

35. (a) M. Agarwal, M. Singh, S. Sharma, M.P.Alam, and C. Chakravorty, J. Phys. Chem. B **114**, 6995 (2010); (b) M. K. Nandi, A. Banerjee, S. Sengupta, S. Sastry, and S. M. Bhattacharyya, J. Chem. Phys. 143, 174504 (2015).

36. R. L Jack and P. Sollich, J. Stat. Mech. **2009**, 11011 (2009).

37. S. Banerjee, R. Biswas, K. Seki, B. Bagchi, J. Chem. Phys. **141,** 124105 (2015).

38. K. Seki and B. Bagchi, J. Chem. Phys. **143**, 194110 (2015).

39. C. M. Newman and D. L. Stein, Ann. l'inst. Henri Poincarè (B) Probab. Stat. 31, 249 (1995), available online at
http://www.numdam.org/item?id=AIHPB_1995__31_1_249_0

40. D. L. Stein and C. M. Newman, AIP Conf. Proc. 1479, 620 (2012).

41. C. M. Newman and D. L. Stein, Phys. Rev. Lett. **72**, 2286 (1994).

42. A. Einstein, Ann. Phys. (Berlin) **322**, 549 (1905).

43. B. Derrida, J. Stat. Phys., **31**, 433 (1983).

44. H. Cordes, S. D. Baranovskii, K. Kohary, P. Thomas, S. Yamasaki, F. Hensel, and J.-H. Wendorff, Phys. Rev. B **63**, 094201 (2001).

45. S. Kirkpatrick, Rev. Mod. Phys. **45**, 574 (1973).

46. S. Havlin, B. L. Trus, and G. H. Weiss, J. Phys. A **19**, L817 (1986).

47. S. Alexander, J. Bernasconi, W. R. Schneier, and R. Orbach, Rev. Mod. Phys. 53, 175 (1981).

48. P. J. H. Denteneer and M. H. Ernst, Phys. Rev. B **29**, 1755 (1984).

49. Th. M. Nieuwenhuizen and M. H. Ernst, Phys. Rev. B **31**, 3518 (1985).

50. (a) M. J. Stephen and R. Kariotis, Phys. Rev. B **26**, 2917 (1982). (b) J. van Polen and H. van Beijeren, Physica A **170**, 247 (1991).

51. O. F. de Alcantara Bonfim and M. Berrondo, J. Phys. A **22**, 4673 (1989).

52. J. Klafter and R. Silbey, Phys. Rev. Lett. **44**, 55 (1980).

53. G. Van der Zwan , J. T. Hynes J. Chem. Phys. **78**, 4174 (1983). **89**, 4181 (1985).

54. S. Okuyama and D. W. Oxtoby, J. Chem. Phys. **84**, 5830 (1985).

55. R. Zwanzig, J. Stat. Phys. **9**, 215 (1973).

56. S. Gosavi, L. L. Chavez, P. A. Jennings and J. N. Onuchic, J. Mol. Biol. **357**, 986 (2006).

57. D. T. Capraro, M. Roy, J. N. Onuchic, and P. A. Jennings, Proc. Natl. Acad. Sci. U. S. A. **105**, 14844 (2008).

58. K. M. Fisher, E. Haglund, J. K. Noel, K. L. Hailey, J. N. Onuchic, and P. A. Jennings, PLoS One **10**, e0144067 (2015).

59. P. C. Whitford and J. N. Onuchic, Curr. Opin. Struct. Biol. **30**, 57 (2015).

60. Abramowitz, M. and Stegun, I. A. (Eds.). *Handbook of Mathematical Functions with Formulas, Graphs, and Mathematical Tables*, ( New York: Dover, 1972).